\documentclass[12pt,preprint]{aastex}

\newcommand{\simle}{\mbox{$\stackrel{<}{_{\sim}}$}}
\newcommand{\simge}{\mbox{$\stackrel{>}{_{\sim}}$}}

\newcommand\msun{\hbox{\,M$_\odot$}}
\newcommand\rsun{\hbox{\,R$_\odot$}}
\newcommand\lsun{\hbox{\,L$_\odot$}}

\shorttitle{G-modes in Rasalhague}
\shortauthors{Monnier et al.}

\begin{document}


\title{Rotationally-modulated g-modes in the rapidly-rotating $\delta$ Scuti star Rasalhague
($\alpha$~Ophiuchi)\thanks{Based on data from the MOST satellite, a Canadian Space Agency mission operated by Dynacon, Inc., the University of Toronto Institute of Aerospace Studies, and the University of British Columbia, with assistance from the University of Vienna, Austria. }}


\author{J.~D.~Monnier\altaffilmark{1}, 
R.H.D. Townsend\altaffilmark{2},
X. Che\altaffilmark{1},
M. Zhao\altaffilmark{3},\\
 T. Kallinger\altaffilmark{4},
J. Matthews\altaffilmark{4},
A. F. J.  Moffat\altaffilmark{5}
}

\altaffiltext{1}{monnier@umich.edu: University of Michigan Astronomy Department, 941 Dennison Bldg, Ann Arbor, MI 48109-1090, USA.}
\altaffiltext{2}{Department of Astronomy, University of Wisconsin-Madison, Sterling Hall, 475 N. Charter Street, Madison, WI 53706, USA}
\altaffiltext{3}{Jet Propulsion Laboratory, Pasadena, CA}
\altaffiltext{4}{University of British Columbia, Vancouver, BC V6T 1Z1, Canada}
\altaffiltext{5}{Universit\'e de Montre\'eal, C.P. 6128, Succ. C-V, QC H3C 3J, Canada}
\email{JDM: monnier@umich.edu}


\begin{abstract}
Despite a century of remarkable progress in understanding stellar
interiors, we know surprisingly little about the inner workings of
stars spinning near their critical limit.  New interferometric imaging
of these so-called ``rapid rotators'' combined with breakthroughs in
asteroseismology promise to lift this veil and probe the strongly
latitude-dependent photospheric characteristics and even reveal the
internal angular momentum distribution of these luminous objects.
Here, we report the first high precision photometry on the low-amplitude
$\delta$~Scuti variable star Rasalhague ($\alpha$~Oph, A5IV,
2.18$\msun$, $\frac{\omega}{\omega_c}\sim0.88$) based on 30~continuous
days of monitoring using the MOST satellite.  We have identified 57$\pm$1
distinct pulsation modes above a stochastic granulation spectrum with
a cutoff of $\sim$26 cycles per day.  Remarkably, we have also
discovered that the fast rotation period of 14.5~hours modulates
low-frequency modes (1-10 day periods) that we identify as a rich
family of g-modes ($|m|$ up to 7).  The spacing of the g-modes is surprisingly linear considering 
Coriolis forces are expected to strongly distort the mode spectrum, suggesting we are
seeing prograde ``equatorial Kelvin'' waves (modes $\ell=m$).  
 We emphasize the unique aspects of
Rasalhague motivating future detailed asteroseismic modeling -- a
source with a precisely measured parallax distance, photospheric
oblateness, latitude temperature structure, and whose low-mass companion provides
an astrometric orbit for precise mass determinations.
\end{abstract}





\section{Introduction}
One of the most enduring questions of astronomy is ``How do stars
work?"  Progress over the last century has led to a robust scientific
framework that explains the physics of internal stellar structure and
also how stars evolve in time.  This framework strives to include
stars of all types, e.g. low metallicity stars of the early universe,
massive stars that will become supernovae, low-mass  brown dwarfs as well as
stars like our Sun.  As a basic rule, our understanding is strongest and
best proven for Sun-like stars and gets shakier and less robust for
stars very much different from the Sun.  This paper will highlight one
critical aspect of stellar structure and evolution, but one that
hardly affects our Sun -- that of stellar rotation.

Stars in general are probably born with significant angular momentum,
but most of them are low-mass (like the Sun) and are ``slowed" down by
their magnetic winds in the early parts of their lives.  On the other
hand, most intermediate- and high-mass stars ($\simge$1.5 solar
masses) have weak magnetic fields and do not live very long, thus are
often observed to be rotating very quickly.  ``Rapid rotation" is
expected to change the star's luminosity and photospheric temperature
distribution \citep{maeder2000} and also strongly modify the observed
surface abundances of various elements \citep{pinsonneault1997}.  For
the most massive stars, rotation will partially determine when the
star becomes supernovae, thus having a crucial impact on how heavy
elements get dispersed into the interstellar medium.  For all its
importance, the effects of rapid rotation are only vaguely understood
and cannot be confidently included in our stellar evolutionary codes.
More observations are critically needed to support the recent
renaissance in theoretical efforts using analytic calculations and
3-dimensional hydrodynamical computer simulations
\citep[e.g.,][]{lee1997,townsend2003a,jackson2005,roxburgh2006,rieutord2006,reese2008}.

Recent breakthroughs in making images of nearby rapidly-rotating stars
using optical interferometry have coincided with new numerical efforts
to simulate the internal stellar structure of these stars
\citep[e.g.,][]{vanbelle2001, souza2003, monnierscience2007,zhao2009}.
Many of the original assumptions
that had been adopted are under renewed scrutiny and simple
assumptions such as rigid-body rotation and early prescriptions for
"gravity darkening" do not seem consistent with new data.

Our group recognized the profound advances possible by combining the
new images of nearby rapid rotators with the asteroseismic constraints
\citep[see also][for general discussion of astereoseismology and interferometry]{cunha2007}
that can be revealed by the precision photometer onboard the MOST
satellite \citep{walker2003}.  We have initially identified two rapidly
rotating stars observable by MOST with extensive published interferometric datasets: Altair and $\alpha$~Oph (Rasalhague).
Since there is no analytic theory that can
predict the effects of rotation on the pulsational modes of a star
spinning at nearly 90\% of breakup, we plan to use the known stellar
and geometric parameters of our sample to constrain the asteroseismic
models with unprecedented power.  Essentially, we know everything
about the external appearance of these stars (size, oblateness
temperature gradients on the surface, viewing angle, rotation periods) from
interferometry and thus we can use asteroseismic signal for deducing
the internal structure without the normally disastrous degeneracies
one confronts.

Here we report on our first results of this project, namely the
oscillation spectrum of $\alpha$~Oph using the MOST observations as
part of the first year NASA Guest Observer program.  $\alpha$~Oph is
classified as an A5IV star \citep{gray2001} with an estimated mass of
2.18$\msun$ and is rotating at 88\% of breakup as judged by the
extreme oblateness of the photosphere observed by interferometry
\citep{zhao2009}.  Interferometric imaging also revealed that our
viewing angle is nearly exactly equator-on, a perspective that
strongly suppresses the photometric amplitudes of $l-|m|$ odd-parity modes,
simplifying mode identifications.  An early K star orbits the primary
star with a period of 7.9 years \citep[based on astrometric and
speckle data, e.g.][]{gatewood2005}, which will allow the masses of
the two stars to be determined eventually to high precision.  Adaptive
optics imaging has recently resolved (Hinkley et al., 2010, private
communication) the two stars and finds the companion to be more than
100$\times$ fainter than the primary at wavelengths of the MOST
instrument and so we have assumed all the pulsation signatures we see
are from the primary.

In this paper we present the full photometric dataset for Rasalhague
along with an updated stellar model based on interferometry data.  We
find evidence for both p-mode and g-mode oscillations, along with a
remarkable rotational modulation signal.  It was beyond the scope of
this paper to develop a mode analysis; however, we will pursue this in
a future paper.

\section{Properties of  $\alpha$ Oph}
\label{alpoph}
The physical properties of $\alpha$~Oph were recently determined
using long-baseline interferometry data from the CHARA Array
\citep{theo2005} using the imaging combiner MIRC
\citep{monnier_spie2004,monnier_spie2006}.  \citet{zhao2009} reported that
$\alpha$~Oph was being viewed within a few degrees of edge-on and 
with a rotation rate $\frac{\omega}{\omega_{\rm crit}}\sim 88$\% of breakup assuming a standard Roche model (point mass gravity, solid-body rotation) and a gravity darkening coefficient of 0.25
\citep{vonzeipel1924a,vonzeipel1924b}.

\begin{figure}[hbt]
\begin{center}
\includegraphics[angle=90,width=6in]{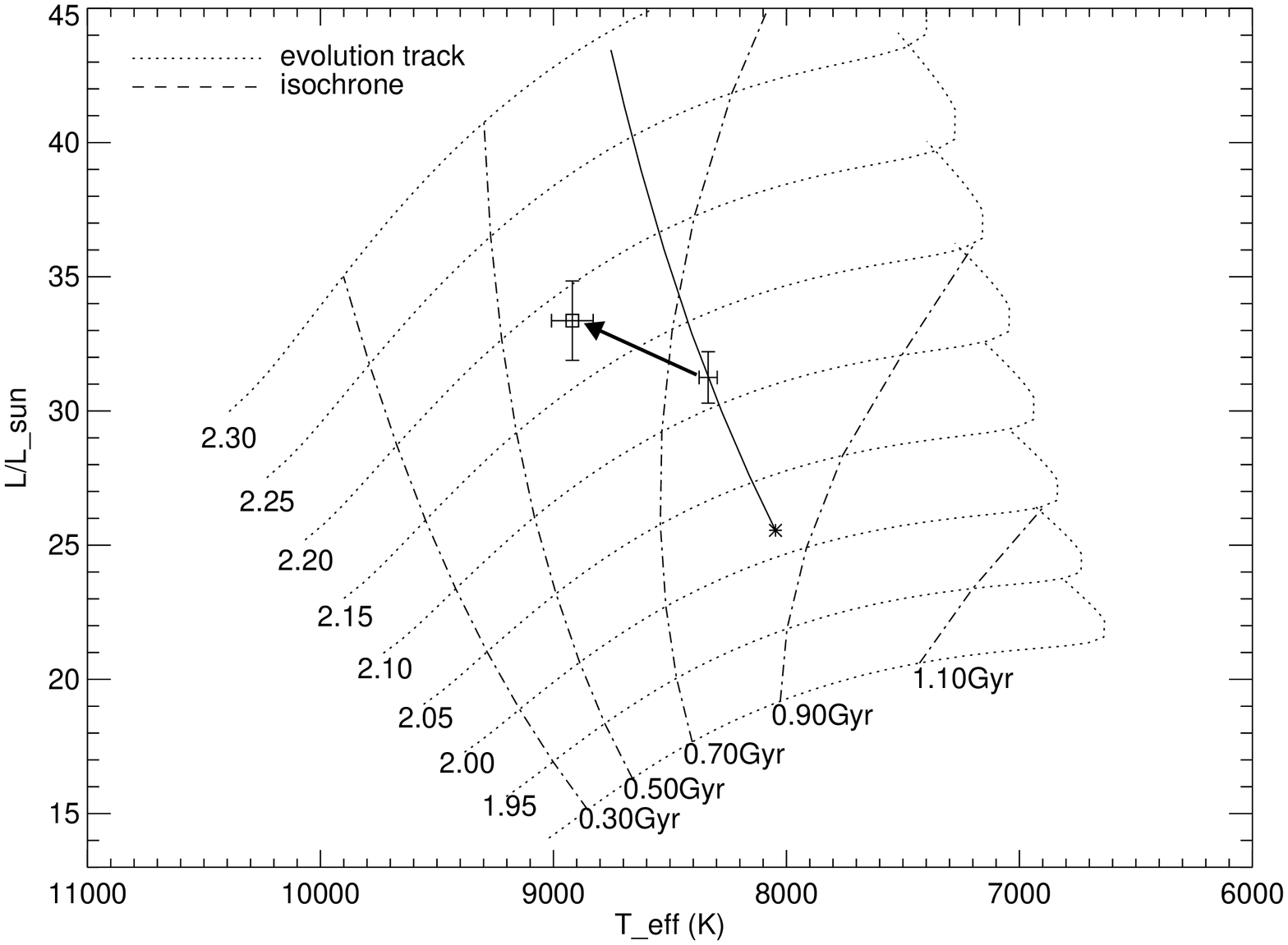}
\figcaption{\footnotesize Based on re-analysis of CHARA-MIRC
  observations  originally presented in
  \citet{zhao2009}, we present a corrected HR diagram for $\alpha$~Oph compared to Y$^{2}$
  isochrones \citep{yi2001,kim2002,demarque2004}.  The asterisk marks the
  ``apparent'' position of $\alpha$~Oph, appearing red and faint due
  to the edge-on viewing angle.  The solid line shows the range of
  possible apparent locations depending on viewing angle with the
  error bars marking the true luminosity and true effective
  temperature. Lastly, the square marks the non-rotating equivalent
  position for $\alpha$~Oph to allow comparison to the  isochrones.
\label{fig_hrdiagram}}
\end{center}
\end{figure}

Since this paper was published, we have slightly improved the
calibration and re-analyzed the existing interferometry data. Most 
importantly, we corrected a methodological error that resulted in
inaccurate age and mass when comparing to non-rotating stellar
evolution tracks.  Figure~\ref{fig_hrdiagram} shows our new estimate of the 
location of $\alpha$~Oph on the Hertzsprung-Russell diagram showing
Y2 isochrones \citep{yi2001,kim2002}.
We have summarized the new model fits and the stellar parameters in Table~\ref{table_alpoph}. Please
see \citet{zhao2009} and Che et al (2010, in preparation) for a detailed description of the model parameters and for details on our fitting methodology.

With the new analysis, we still find $\alpha$~Oph to be rotating 88\%
of breakup (rotation rate of 1.65 c/d) and being viewed within a few
degrees of edge-on.  However, the derived mass is now a bit higher --
2.18\msun -- a result of the improved isochrone analysis. 
We will adopt these parameters in discussions throughout the rest of
the paper.  Note that the error bars in the table rely on fixing the
gravity darkening coefficient $\beta$ to 0.25.  Evidence is emerging
(Xiao Che, private communication) that $\beta$ might be substantially
lower than this, which could affect the derived stellar parameters.

\section{Observations}
\label{observations}

\subsection{Basic Description}
The photometry presented here was obtained by the Microvariability and Oscillations of Stars (MOST) 
satellite. MOST houses a CCD photometer fed by a 15-cm Maksutov telescope through
a custom broadband optical filter (350 - 750nm). The satellite's
Sun-synchronous 820km polar orbit (period $=$ 101.4min $=$ 14.20
cycles/day) enables uninterrupted observations of stars in its
Continuous Viewing Zone (-18\arcdeg $<$ dec $<$ +36\arcdeg) for up to
8 weeks. A pre-launch summary of the mission is given by
\citet{walker2003} and on-orbit science operations are described by
\citet{matthews2004}.

The MOST photometry of the $\alpha$~Ophiuchi binary system covers 30
days during 2009 May 27 to 2009 Jun 26, and was obtained in Fabry
Imaging mode, where the telescope pupil, illuminated by the star, is
projected onto the CCD by a Fabry microlens as an annulus covering
about 1300 pixels. The data were reduced using the background
decorrelation technique described by \citet{reegen2006}. The reduction
includes filtering for cosmic particle impacts, correction of the
stray light background modulated with satellite orbital parameters,
filtering of long-term trends, and removal of obvious outliers.

The final heliocentric corrected time series provided by the MOST
instrument team contained 81018 measurements with an exposure time and
sampling of 30 seconds (based on thirty 1-sec exposures stacked
onboard the spacecraft). The datastream was remarkably consistent and
continuous, being on-target with data for about 94\% of the $\alpha$~Oph
observation. Instrument problems did cause six data gaps in excess of
30 minutes, with the longest being 5 hours. 

\subsection{Data reduction}
\begin{figure}[hbt]
\begin{center}
\includegraphics[angle=90,width=6in]{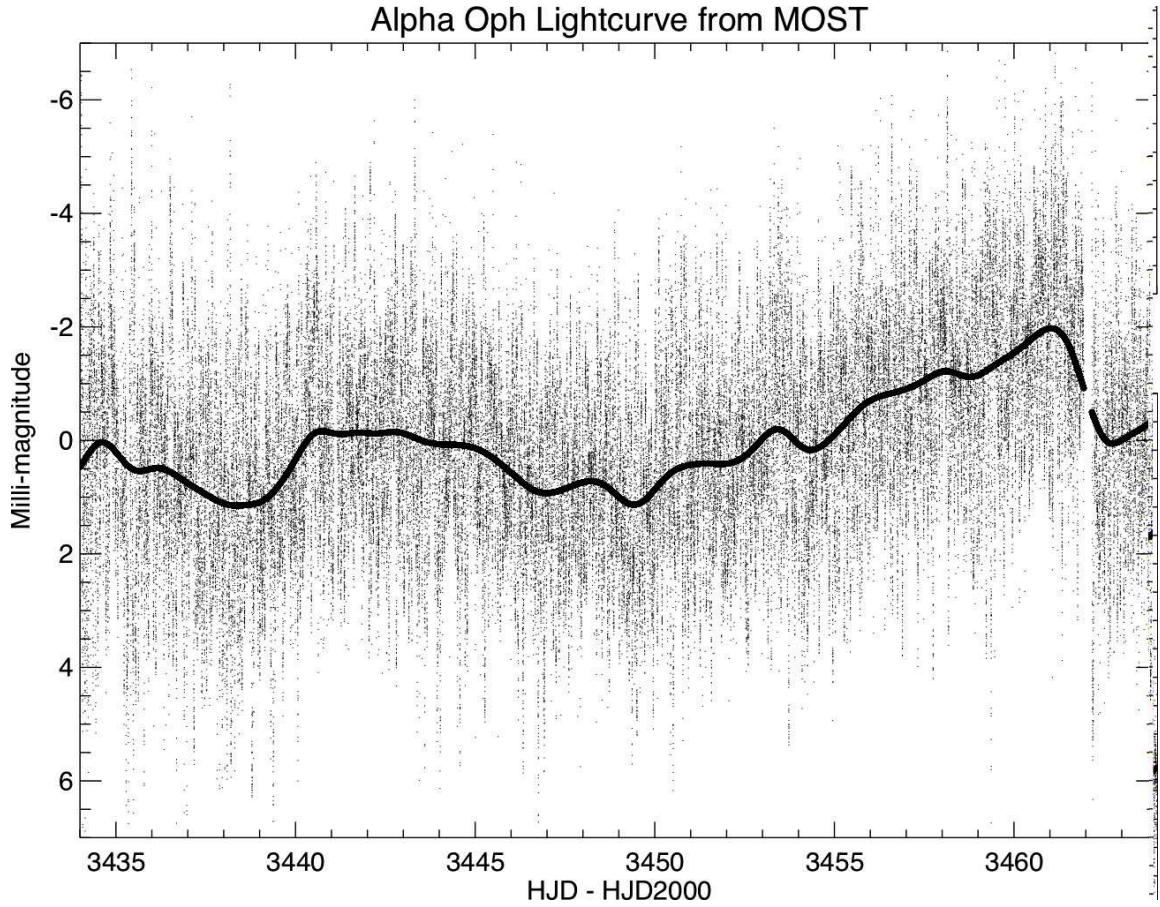}
\figcaption{\footnotesize This figure shows the MOST photometry of $\alpha$~Oph (Rasalhague) in an early stage of data processing.  Each tiny point is a measurement over a 30-second period and the solid line shows the 1-day moving average.  
\label{fig_lightcurve1}}
\end{center}
\end{figure}

The first step our team applied in our data analysis was to remove local 5-sigma outliers
from the data.  Because a similar procedure had already been applied
by the MOST instrument team, this step had only a minor effect --
reducing the dataset to 80002 independent data points, a data volume
reduction of only 1.3\%.  These flux points are presented in
Figure~\ref{fig_lightcurve1} in their entirety.

The next step  was to remove the harmonics of the 14.2 c/d
orbital period. Similar to the method described by
\citet{walker2005}, we used a moving window of 10~days to create a high
signal-to-noise mean light curve folded at the orbital period. We then
remove this component from the original light curve.  The long
averaging window ensures we only remove frequencies within a few (Nyquist)
resolution elements of the harmonics.  We have inspected data both
with and without this correction and found the only significant effect
was to remove the harmonics from our final power spectra.  

Lastly, we ``detrended'' the lightcurve in order to filter out slow
variations on long time scales.  The ``slow trend'' light curve was
constructed by fitting a line segment to the flux measurements (after
removal of orbital modulation) within a moving 1-day window. The slow
trends were subtracted from the lightcurve resulting in a final data set
on which subsequent Fourier analysis was performed.  The slow trend
light curve can be seen in Figure~\ref{fig_lightcurve1}.  At our
request, spacecraft telemetry was inspected to see if any camera or
telescope diagnostic showed variations in sync with the slow
variations seen here -- none were found.

\begin{figure}[hbt]
\begin{center}
\includegraphics[angle=90,width=6in]{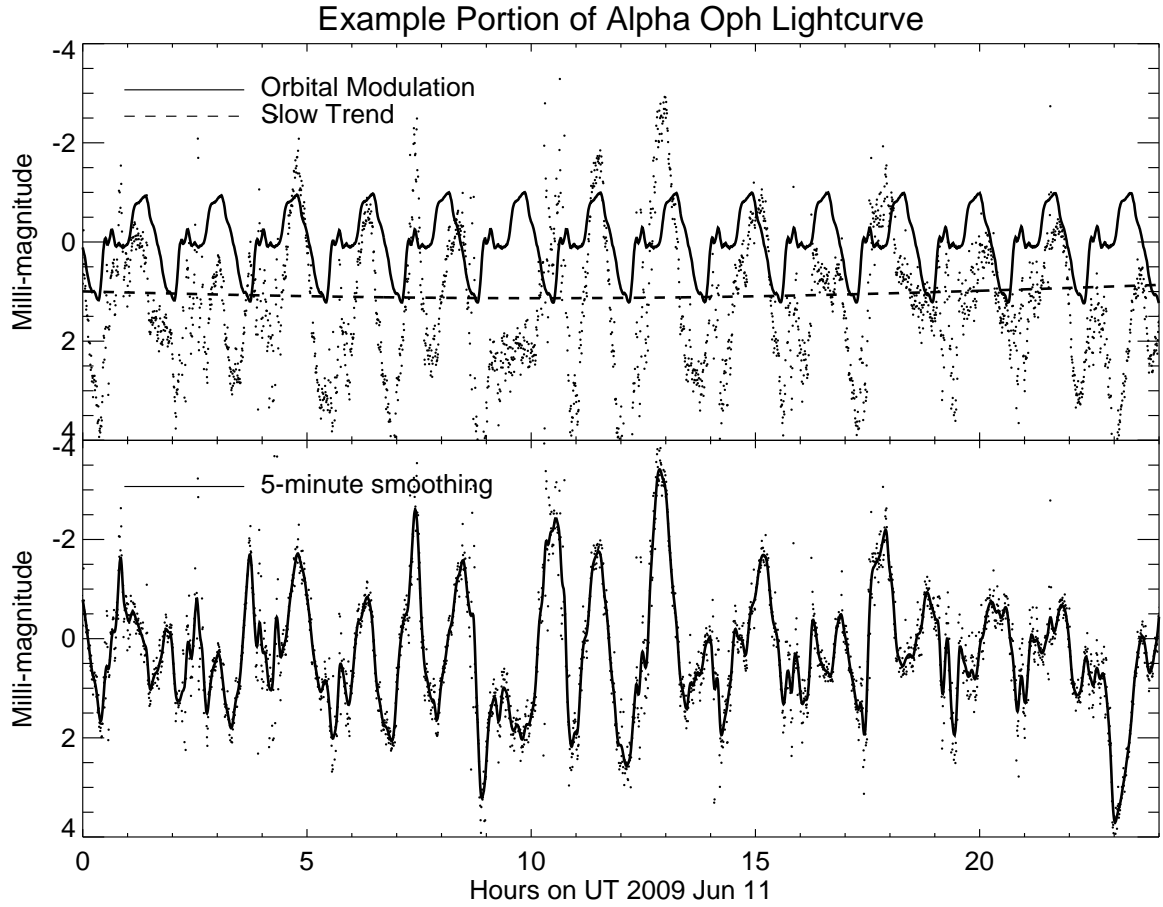}
\figcaption{\footnotesize This figure shows each step in the data reduction for the lightcurve of $\alpha$~Oph (Rasalhague).  The top panel shows the original data for an example 1-day period along with the estimated orbital modulation and slow trend components.  The bottom panel shows the data after removing the orbital modulation and the slow trend -- this final reduced light curve was used for Fourier analysis in this paper.
\label{fig_lightcurve2}}
\end{center}
\end{figure}

We have graphically summarized the steps in our data reduction in
Figure~\ref{fig_lightcurve2}. Here we have zoomed up on a single day in
our time series and show explicitly the original data, our estimate
of the residual orbital modulation, and the slow trend.  We also show
the final corrected light curve with and without a 5-minute averaging
window to demonstrate the overal quality of our data.

After more than 7 years in orbit, the MOST Team has a very robust
understanding of the performance of the instrument, especially the
point-to-point precision of the photometry as a function of target
magnitude and the noise levels at various frequency ranges in
Fourier space.  Many of the pulsating variable stars detected in
the MOST Guide Star sample (in the magnitude ranges 6 $<$ V $<$ 10) had
a ``discovery amplitude" (i.e., amplitude of the largest oscillation
signal) of only about 1 - 2 mmag, with many other oscillation peaks
ten times smaller and a noise level in Fourier space 50 times lower
\citep[e.g., the richly multi-periodic delta Scuti star HD 209775][]{matthews2007}.  Some examples of stars which characterise the photometric
precision of the MOST instrument are: HD 146490 \citep[a guide star used
as a calibration by][; see their Figure 1]{kallinger2008b}, with
V $=$ 7.2 and point-to-point scatter of 0.12 mmag; tau Bootis (V $=$ 4.5),
an exoplanet-hosting star which shows clear modulation at the planet's
orbital period with a peak-to-peak amplitude of 2 mmag 
\citep{walker2008} for which its mean instrumental magnitude is repeatable to
within 1 mmag in observations spaced several years apart; and Procyon
(V $=$ 0.35), with point-to-point precision of 0.14 mmag.  In these
cases, the noise floor in the Fourier spectrum is well defined and
at a level of 0.01 mmag or less.

Another MOST target, epsilon Oph \citep{barban2007,kallinger2008a}, a cool giant (V $=$ 3.24), is stable to 0.35 mmag at frequencies
below 0.1 c/d.  At these low frequencies, non-periodic noise intrinsic
to the star is usually associated with granulation.  The MOST Team
has never seen instrumental artifacts with the amplitudes and coherence
of the variations evident at low frequencies in the MOST light curve of
alpha Oph presented here.  For the results in this paper, the low-frequency
signals have no direct implications, so we do not address them in further
detail.  However, they are likely intrinsic to the star and hence do
not raise any alarms of possible unrecognised photometric artifacts
at other frequency ranges in the alpha Oph time series.

\section{Analysis}
\label{analysis}

\subsection{Amplitude Spectrum}

We have carried out an exhaustive Fourier analysis of the MOST
photometry.  Although the MOST data are remarkably complete and
evenly-spaced compared to ground-based data, we nonetheless
employed a numerical Fourier Transform followed by a ``CLEAN''ing step
based on the algorithm of \citet{roberts1987} in order to remove the small
sidelobes from the strongest peaks in the power spectrum that arise
due to the gaps in the temporal sampling.

Figure~\ref{fig3} shows the amplitude spectrum of our photometry in
units of milli-magnitude for the range 0 to 64~c/d. The power for
frequencies below $\sim$1~c/d was suppressed by the detrending step
described in the previous section.  In fact, without detrending, nearly all
the frequencies below $\sim$0.75~c/d show significant power above
background noise levels. We note that all frequencies up to 1440~c/d were
inspected and frequencies above 64~c/d are not shown here because no
statistically significant peaks were found.

\subsection{Non-white ``noise''}
\label{nonwhitenoise}
It was clear from the power spectrum that the background noise level
is higher for frequencies below 30~c/d compared to higher frequencies.
In order to properly estimate the statistical significance of a
``peak'' it is crucial to establish the level of the
frequency-dependent noise background.  We have used a median filter with a 2~c/d
window in order to crudely estimate the background noise power for a
given frequency, and this background level is included in
Figure~\ref{fig3}.  We will discuss the physical nature of this
non-white noise in \S\ref{granulation}; for now we simply treat it as
a noise source that can produce spurious spikes in the amplitude spectrum.

\subsection{Identification of Distinct Modes}

\begin{figure}[p]
\begin{center}
\includegraphics[width=6in]{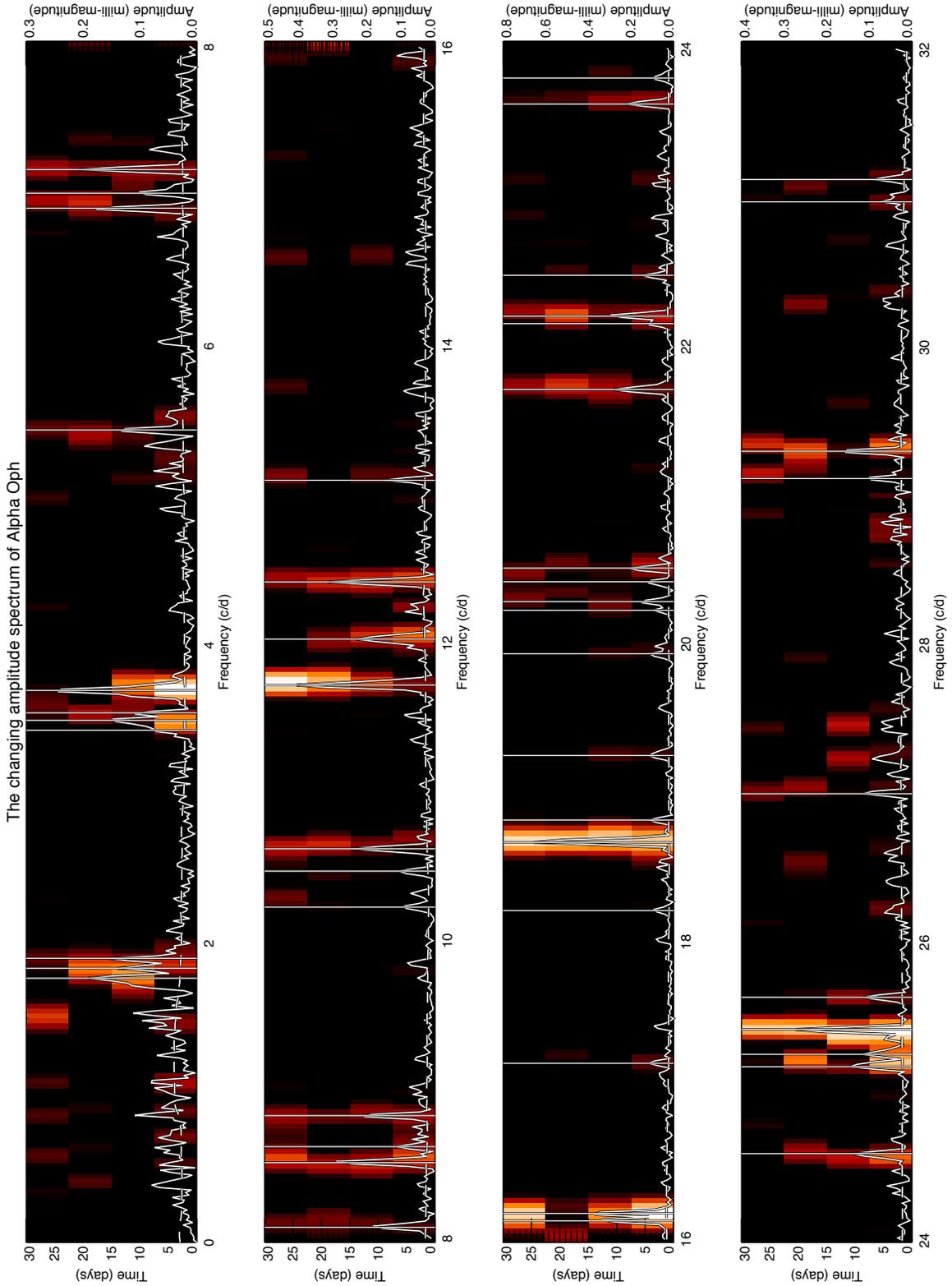}
\figcaption{\footnotesize Here we present our Fourier Analysis of the lightcurve of $\alpha$~Oph.  The solid line shows the amplitude spectrum based on the full 30-day data set, with the amplitude labeled on the right-hand axis.  Our estimate of the median background noise level is shown with a dashed line and 3.5-$\sigma$ peaks are shown with vertical lines.   Here we also show the week-by-week temporal variability as a background image of the power spectrum, shown here in logscale.  There were no statically-significant peaks discovered above 50 c/d.
\label{fig3}}
\end{center}
\end{figure}

Armed with the frequency-dependent background noise spectrum estimated
in the last subsection, we can now place confidence limits on a given
amplitude peak being a ``real'' distinct pulsation mode rather than
having arisen by chance due to background fluctuations.  We have
chosen a statistical measure such that we only expect one of our
identified modes in the range between 0.5 and 64 c/d to be a false
positive.  Since there are
$\sim$1800~independent frequencies that can be probed in our data in
this spectral range, we chose only peaks in the power spectrum above
3.5-$\sigma$ confidence, a criteria that should produce roughly 1 false positive
assuming pure gaussian random noise.  The 57~modes that survived this
analysis are listed in Table~\ref{table_modes} and are plotted in
Figure~\ref{fig3}. We also confirmed the statistical significance of our peaks through a bootstrap analysis and we note that we would have erroneously identified $>$500 distinct modes if we had ignored the frequency-dependence of the noise background.

The strongest modes are 18.668 c/d (half-amplitude 0.66~mmag), the
closely spaced pair at 16.124/16.174 c/d (0.35/0.41 mmag), and at
11.72 c/d (0.41~mmag).  These frequencies are typical for the dominant
p-mode oscillations in $\delta$~Scuti stars and bear some resemblance
to the dominant frequencies observed for another rapid rotator Altair
\citep{buzasi2005}, which had dominant peaks at approximately 15.7,
20.8, and 26.0 c/d.  The highest frequency mode we have detected with
high confidence is at 48.35 c/d (0.036 mmag).

Inspection of Figure~\ref{fig3} reveals many additional peaks that
appear real -- more than can be attributed to random fluctuations of
the background.  However, we have chosen not to report these peaks in
Table~\ref{table_modes} in order to maintain a highly pure set of
frequencies with minimal false positives.  In other words, we
understand that many of the other peaks are real pulsational
frequencies but are also sure that some of them are contaminated by
background fluctuations.  Future mode analysis might allow some of the
2- or 3-$\sigma$ peaks to become bona-fide pulsation frequencies, but
we have been conservative in our mode identifications for this paper.

In addition, inspection of the amplitude spectrum at lower frequencies
easily reveals increased power at harmonics of 1.7 c/d, close to the
expected rotational frequency (1.65 c/d, see \S\ref{alpoph}).  We will
discuss the physical origin of these frequencies in the next section.

\subsection{Time Variability}
With such a long dataset of high quality, we investigated the
stability of the detected modes.  Figure~\ref{fig3} contains also the
power spectrum split into four 1-week chunks and is depicted as a 2-D
background image in each panel.  One can see the lower frequency
modes, $<$10 c/d, tend to be variable -- strong in some weeks and not
present in others.  \citet{degroote2009} found amplitude variability
in modes in some higher-mass objects with liftimes from one to a few
days, substantially shorter than those here which tend to be
approximately one week.  Most of the higher frequency modes are more
stable, although not all. For instance the 16.1~c/d dominant mode
shows amplitude varying by a large amount, consistent with two equal
modes with a frequency difference just barely resolved by our 30~day
run.  The lifetime of modes is another clue to their physical origin
and will be discussed in the next section.

\section{Discussion}
\label{discussion}

Here, we wish to discuss three aspects of this dataset: the detection of granulation noise $<$30 c/d, the  p-mode spectrum, and lastly the discovery of g-modes and their rotational modulation.

\subsection{Granulation}
\label{granulation}

\citet{poretti2009} observed the $\delta$~Scuti star HD~50844 using
COROT, identifying $>$1000 statistically-significant  pulsational modes using standard
Fourier analysis.  \citet{kallinger2010} argued it was physically unreasonable 
 to have so many distinct modes and interpreted the higher
power for frequencies lower than $\sim$30 c/d as due to granulation
noise.  Indeed, these authors presented a compelling  correlation of the
granulation cutoff frequency with stellar mass and radius.
We subscribe to the interpretation of Kallinger \& Matthews, choosing
to treat the higher average Fourier power at the lower
frequencies as a kind of non-white background noise (see previous
section \S\ref{nonwhitenoise}.  

In order to better estimate the granulation cutoff frequency we have
plotted the median-filtered power (2 c/d window) in
Figure~\ref{fig_noise}.  We identified an obvious artifact from the
orbital harmonics at 42.6 c/d, and 56.8 c/d and likely present at 14.2
c/d and 28.4 c/d, probably due to imperfect background subtraction of
scattered light.
By modeling the harmonics beyond 35 c/d, we have crudely estimated
the contamination at lower frequencies based on the isolated features
at 42.6 and 56.8 c/d.  Figure~\ref{fig_noise} shows our ad hoc model
for the contributions from the orbital harmonics and also includes our corrected
granulation noise spectrum.  Here, the granulation cutoff frequency can be estimated to be 
26$\pm$2 c/d.  

As explained in \citet{kallinger2010}, the granulation cutoff
frequency should be related to the inverse of the sound crossing time
of one pressure scale height in the outer convective layer.  This
leads to a dependence of $\nu_{\rm granulation} \propto
\frac{M}{R^2~T^\frac{1}{2}}$.  It is beyond the scope of this paper to
compare our result with their more elaborate power-law fitting method;
however, we can compare our measured granulation cutoff with that
predicted by the empirical relation discovered by \citet[][see their figure
  3]{kallinger2010}.  Using a mass of 2.18\msun, effective radius of
2.7\rsun and $T_{\rm eff}=8300K$ for $\alpha$~Oph, we find that
$M~R^{-2}~T^{-0.5} \sim$0.25 (in solar units), leading to an expected
$\nu_{\rm granulation}\sim 300 \mu$Hz$=26$c/d, consistent with our observations.

\subsection{Rotationally-modulated g-modes}
\label{gmodes}
Following treatment of \citet{buzasi2005}, we can estimate the frequency of the fundamental radial (p-) mode  
using the relation 

\begin{equation}
P \sqrt{\rho/\rho_\odot} = Q
\end{equation}

For $Q=0.033~$days \citep{breger1979} and using the stellar parameters
from Table~\ref{table_alpoph}, we find $f_{\rm fun}\sim$10.1~c/d
(compared to 15.6~c/d for the 1.79$\msun$ star Altair).  Thus, we will
begin by assuming $p$-modes  all have frequencies above or near to
this cutoff.  A full analysis of the p-modes with tentative mode
identifications will be the subject of a future paper. Here, we focus
primarily on the unexpected discovery of rotationally-modulated
g-modes in $\alpha$~Oph.

We showed in the last section (see also Figure~\ref{fig_lightcurve1})
that there was a $\sim$1~mmag slow (week- timescale) variation seen in
the photometry.  We also have identified 15 strong modes with
frequencies smaller than the fundamental radial mode that cluster
around harmonics of 1.7 c/d, close to the rotation frequency.  This
distinctive structuring suggests we are witnessing the rotational
modulation of modes whose frequency in the co-rotating frame is small
in comparison to the rotation frequency.

The tabulated frequencies were extracted directly from peaks in the
CLEANed Fourier Transform and we assigned frequency errors based
simply on the total length of the dataset.  We realize that higher
precision on the frequency localization of pure modes can sometimes be
achieved through a multi-component least-squares fit
\citep[e.g.][]{walker2005}, however this procedure has convergence
problems for very closed spaced frequencies as encountered here.  The
errors on amplitudes were assigned based on the local background noise
as estimated in \S 4.2.  We caution that the formal amplitude errors
may be underestimates due to non-stationary statistics of the
background or due to amplitude variations in the modes themselves.
The full time series is available upon request to researchers
interested in exploring alternative analysis schemes, but our
straightforward approach adopted here leads to conservative errors in the
frequencies and is sufficient for the analysis presented here.

Gravity (g-) modes are expected to have long periods compared to
p-modes, although it has been difficult to estimate this theoretically
for rapid rotators.  Indeed, our amplitude spectrum is consistent with
a complex and time-variable set of g-modes with co-rotating
frequencies $\simle$0.1~c/d, corresponding to approximately 10-day
periods.  Normally the properties of these modes are quite difficult
to measure due to the long time frames needed to see the modes travel
around the star.  However, in our case, the $\sim$1.65 c/d rotational
rate of this rapid rotator leads to strong rotational modulation.
Note also that modes of North-South asymmetry are invisible to us
while the symmetric $l-|m|$ even-parity modes are strongly modulated due to
our near equator-on viewing angle.

Specifically we see small clusters of modes around frequencies
$\sim$1.8, 3.5, 5.4, 7.0, 8.6, 10.4, 12.0 c/d.  While some of the
higher frequencies here might be p-modes, it seems that the strong
periodicity at about 1.7 c/d suggests that each cluster represents a
new non-radial family of modes with increasing $m$ values.  Stellar
rotation causes lightcurve variations with frequencies proportional to
$m f_{\rm rot}$, since the $m$ parameter represents the number of
nodes around the equator. With this interpretation, we are seeing
non-radial modes corresponding to $m=1,2,3,4,5$ and possible
$m=6,7$. We believe this is the first time such a rich set of g-modes has been
detected and partially identified around a rapid rotator, although perhaps a related phenomenon may be at
work around the late Be star HD~50209 \citep{diago2009}.

The simultaneous appearance of g-modes and p-modes would make $\alpha$~Oph a ``hybrid''
$\gamma$~Dor-$\delta$~Sct pulsator, a class recently explored by \citet{grigahcene2010} using first Kepler data. Indeed, these authors present a frequency spectrum of KIC9775454 that bears some similarity to the data presented here for
$\alpha$~Oph, although our frequency spectrum
does not show the characteristic ``gap'' between 5 and 10 c/d. 
Further comparison with this work is not possible until more is known about the rotational properties of the Kepler sample, especially since many hybrid pulsators are known to be slowly-rotating Am stars.

To further investigate the periodic spacing of the g-modes of $\alpha$~Oph, 
we searched for families of modes that
follow a linear frequency relationship of the form
\begin{equation}
f = f_{0} + m f_{1}
\end{equation}
for integer $m$. We found two families, each comprising more than
three modes, that can be well represented by this formula. Family A
contains 5 modes; assuming the lowest-frequency mode corresponds to
$m=1$, a least-squares fit gives $f_{0} = 0.073 \pm 0.017$\,{\rm c/d}
and $f_{1} = 1.7095 \pm 0.0002\,{\rm c/d}$. Family B contains 4 modes,
with a fit $f_{0} = 0.219 \pm 0.017\,{\rm c/d}$ and $f_{1} = 1.7424
\pm 0.0002\,{\rm c/d}$. These modes are noted in Table~\ref{table_modes} and included
in Figure~\ref{fig_modes} along with
the observed amplitude spectrum for $f < 16\,{\rm c/d}$. There are
other solutions that fit three modes or fewer, but in this treatment
we consider just Family A and B, since the larger number of modes mean
the fit is more reliable and less likely due to chance superpositions.

There are a number of possible interpretations for the linear
frequency relationship exhibited by the two mode families, which we
shall consider in turn. To first order in the rotation frequency
$f_{\rm rot}$, the observed frequency of a mode with radial order $n$,
harmonic degree $\ell$ and azimuthal order $m$ is given by the
well-known \citet{ledoux1951} formula,
\begin{equation} \label{eqn:f-ledoux}
 f = f_{n,\ell} + m f_{\rm rot} (1 - C_{n,\ell}).
\end{equation}
Here, $f_{n,\ell}$ is the frequency the mode would have in the absence
of rotation, and is independent of $m$. The parentheses combine
together the effects of the Doppler shift (in transforming from
co-rotating to inertial reference frame) and the Coriolis force ---
the latter represented by the $C_{n,\ell}$ term, which again is
independent of $m$. For high-order g modes, $C_{n,\ell}$ approaches
$[\ell(\ell+1)]^{-1}$ \citep[e.g.,][]{aerts2010}.

If we make the identifications $f_{0} = f_{n,\ell}$ and $f_{1} =
f_{\rm rot} (1 - C_{n, \ell})$, then the Ledoux formula is apparently
able to explain the linear frequency relationship of the two families,
assuming each comprises modes having the same $n$ and $\ell$ but
differing $m$. However, there are strong arguments against this
conclusion. For both families the mode frequencies in the co-rotating
frame,
\begin{equation} \label{eqn:f_c}
f_{\rm c} \equiv f - m f_{\rm rot} 
\end{equation}
are small (on the order of $0.1-0.3\,{\rm c/d}$) compared to the
rotation frequency $f_{\rm rot} = 1.65\,{\rm c/d}$ (Table~\ref{table_alpoph}). The spin
parameter $\nu \equiv 2 |f_{\rm rot}/f_{\rm c}|$ of the modes must
therefore be much larger than unity, implying that their dynamics are
dominated by the Coriolis force \citep[see, e.g.][]{townsend2003a}. This
represents a fundamental inconsistency with the use of the first-order
Ledoux formula (\ref{eqn:f-ledoux}), which is valid only for $\nu
\lesssim 1$.

A possible way around this inconsistency is to relax the assumption
that lowest-frequency mode in each family corresponds to azumithal
order $m=1$. If we instead assume $m \geq 3$ for these modes, then by
eqn.~(\ref{eqn:f_c})\footnote{We note that negative $f_{\rm c}$ values in
  this equation indicate modes whose phase propagation in the
  co-rotating frame is retrograde.} we can achieve $\nu \lesssim
1$. However, this `fix' itself runs into difficulty when we note that
for both mode families $f_{1} > f_{\rm rot}$, implying that the
Coriolis coefficients $C_{n,\ell}$ are negative. This is incompatible
with the above-mentioned limit $C_{n,\ell} \approx [\ell(\ell+1)]^{-1}
> 0$ (indeed, we are not aware of any physically plausible pulsation
model that predicts negative Coriolis coefficients).

Accordingly, we are led to abandon the Ledoux formula as an
explanation for the linear frequency relationship of the two mode
families. In searching for alternative interpretations, we note that
the small magnitude of $f_{0}$ suggests we may be observing
7\emph{dispersion-free} low-frequency modes. In rotating stars there
are two classes of mode that are dispersion-free: equatorial Kelvin
modes (which are prograde) and Rossby modes (which are
retrograde). Rossby modes tend not to generate light variations be-
cause they are incompressible, so let us put them aside for the moment
(although not forget about them completely).

Focusing therefore on the Kelvin modes, these are prograde sectoral
($\ell = m$) g-modes which are confined to an equatorial waveguide by
the action of the Coriolis force \citep[see][]{townsend2003a}. A defining
characteristic of these modes is that their frequencies are directly
proportional to $m$. At high radial orders, these frequencies in the
co-rotating frame can be approximated by
\begin{equation}
f_{\rm c} \approx \frac{m I_{\rm g}}{2\pi^{2}n};
\end{equation}
here, following \citet{mullan1989}, we define
\begin{equation}
I_{\rm g} = \frac{1}{\sqrt{2} \pi^{2}} \int_{0} \frac{N}{r} \,{\rm d}r,
\end{equation}
with $N$ is the Brunt-V\"ais\"al\"a frequency. The corresponding
observed frequencies are then given by
\begin{equation}
f \approx m \left( \frac{I_{\rm g}}{2 \pi^{2} n} + f_{\rm rot} \right).
\end{equation}
For a group of modes all having the same $n$, this expression predicts
a linear frequency relationship \emph{with zero intercept}. If $n$
varies within the group, however, then there will be some scatter
about a linear relationship, with a relative amplitude of $\sim \Delta
n/n$.

To examine whether the two mode families could be Kelvin modes, we
re-fit their frequencies assuming a zero intercept (again identifying
the lowest-frequency mode as $m=1$). This gives $f_{c}/m = 0.07 \pm
0.03\,{\rm c/d}$ for Family A, and $f_{c}/m = 0.10 \pm 0.03\,{\rm
  c/d}$ for Family B. To interpret these values, we assume a
relationship $I_{\rm g} \approx 10^{-3}
\sqrt{(M/M_{\odot})/(R/R_{\odot})^{3}}\,{\rm Hz}$, derived from the
$n=3$ polytropic model considered by \citet{mullan1989}. Using the stellar
parameters from Table~\ref{table_alpoph}, this yields $I_{\rm g} \approx 34\,{\rm
  c/d}$. Hence, the approximate radial orders of the modes are derived
as $n \approx 24$ (Family A) and $n \approx 17$ (Family B), with
corresponding scatter (based on the quoted uncertainties in the
$f_{c}/m$ fits) of $\Delta n \approx 10$ and $\Delta n \approx 5$,
respectively.

These values are consistent with the typical radial orders of unstable
g modes observed in the more-massive Slowly Pulsating B (SPB) stars
\citep[e.g.,][]{cameron2008,pamyatnykh1999}. This lends strong support to the
identification of the two mode families as high-order ($n \approx 17$
and $n \approx 24$) g modes transformed by the Coriolis force into
equatorial Kelvin modes.

As a last comment, according to Figure~\ref{fig3}, most of the g-modes
seem to be time-variable within our one-month time frame, in contrast
to the p-modes whose amplitudes are more typically constant within the
errors.  The g-modes tend to change on a time scale similar to the
intrinsic mode period ($\sim$10 days) or a bit longer.  This could be
due to superposition of many closely spaced modes in what appears to
be indeed a complex spectra.  Alternatively, many excitable modes may be
transferring energy between themselves, leading to a dynamic and
time-variable amplitude spectrum.  The ability to see so many
independent $\ell$ and $m$ modes by using strong rotational modulation is
invaluable to deriving strong constraints on the mode properties up
to high order (here up to $m=7$) and promises to reveal interior stellar 
structure with additional analysis.

\section{Conclusions}
This first long stare by MOST at a rapidly rotating A star has led to
number of a new results.  We detect for the first time a rich p-mode
spectrum consistent with low-amplitude $\delta$-Scuti pulsations, and
measure a granulation spectrum below 26$\pm$2~c/d.  In total, we have
identified 57$\pm$1 distinct modes below 50 c/d including a complex set
of low-frequency modes that we identify as rotationally-modulated
g-modes with (co-rotating) frequencies $\sim$0.1 c/d.  A mode analysis
revealed linear relationships between the spacings of g-modes up to
$m=7$, an unexpected result for a star rotating at $\sim$90\% of
breakup.  This periodicity can be explained as due to dispersion-free
equatorial Kelvin waves (prograde $l=m$ modes) although some inconsistencies in
our analysis demand follow-up study. Lastly, the long time-base has
allowed us to study the mode lifetime, finding that most p-modes are
stable while g-modes appear to live only a few times their intrinsic
(co-rotating) periods.

Understanding the effect of rapid rotation on stellar interiors is
crucial to developing reliable models for massive star evolution in
general.  $\alpha$~Oph is emerging as a crucial prototype object for
challenging our models and to spur observational and theoretical
progress.  Additional work is planned using adaptive optics to
determine the mass of the star by measuring the 8-year visual orbit
more precisely, using visible and infrared interferometry to strictly constrain possible
differential rotation and gravity darkening laws, and using
asteroseismology to follow-up on the new discoveries outlined here.

\acknowledgments {  We thank C. Matzner for discussion and insights.
  TK is supported by the Canadian Space Agency and the  Austrian Science Fund.
  AFJM is grateful for financial aid from NSERC (Canada) and FQRNT (Qu\'ebec).
  We acknowledge support from the NASA MOST guest observer program
  NNX09AH29G and NSF AST-0707927. This research has made use of the SIMBAD database, operated at CDS, 
  Strasbourg, France, and  NASA's Astrophysics Data System (ADS) Bibliographic Services.'' 

{\it Facility:} \facility{MOST}, \facility{CHARA (MIRC)}
}

\bibliographystyle{apj}
\bibliography{apj-jour,most,v892tau,RS_Oph,Review2,Review,IONIC3,iKeck,KeckIOTA}

\clearpage

\begin{deluxetable}{lcc}
\tabletypesize{\scriptsize}
\tablecaption{Best-fit and physical parameters of $\alpha$~Oph from
CHARA-MIRC interferometry \label{table_alpoph}}
\tablewidth{0pt}
\tablehead{ 
 \colhead{Model parameters} &
}
\startdata
Inclination (degs)          &  87.5  $\pm$ 0.6             \\
Position Angle (degs)  &  -53.5 $\pm$ 1.7             \\
T$_{pol}$ (K)                 &  9384   $\pm$ 154              \\
R$_{pol}$ (mas)           &  0.757  $\pm$ 0.004           \\
$\omega$                     &  0.880  $\pm$ 0.026            \\
$\beta$                          &  0.25 (fixed)                        \\
\hline
\hline
\colhead{Derived physical parameters} & \\
\hline
T$_{equ}$  (K)                 &  7569   $\pm$ 124              \\
R$_{equ}$  ($\rsun$)    &  2.858  $\pm$ 0.015           \\
R$_{pol}$ ($\rsun$)    &   2.388 $\pm$ 0.013	              \\
True T$_{eff} (K)$ & 8336 $\pm$ 39 \\  	
True Luminosity (\lsun) & 31.3 $\pm$ 0.96 \\
Apparent T$_{eff} (K)$ & 8047  \\
Apparent Luminosity (\lsun) & 25.6 \\
V Magnitude\tablenotemark{a}     & 2.087     \\
H Magnitude\tablenotemark{b}     & 1.709       \\
v $\sin i$ (km/s)                                  & 239   $\pm$ 12    \\
Rotation rate (cycles/day)                   & 1.65 $\pm$ 0.04\\
Mass (\msun)\tablenotemark{c} & 2.18 $\pm$ 0.02 \\
Age (Gyrs)\tablenotemark{c} & 0.60 $\pm$ 0.02  \\
\hline
\hline
\colhead{$\chi^2_{\nu}$ of various data} & \\
\hline
Total $\chi^2_{\nu}$                                   & 0.70       \\
CP $\chi^2_{\nu}$                                      & 1.08       \\
Vis$^2$ $\chi^2_{\nu}$                             & 0.86        \\
T3amp $\chi^2_{\nu}$                               & 0.16        \\
\hline
\hline
\colhead{Physical parameters from literature} & \\
\hline
$[Fe/H]$\tablenotemark{d}  & -0.16 \\
Distance (pc)\tablenotemark{e} & 14.68\\
\enddata
\tablenotetext{a}{V magnitude from literature: 2.086 $\pm$ 0.003 \citep{perryman1997}.}
\tablenotetext{b}{H magnitude from literature: 1.72  $\pm$ 0.18  \citep{cutri2003}.}
\tablenotetext{c}{Based on the $Y^2$ stellar evolution model \citep{demarque2004}.}
\tablenotetext{d}{\citet{erspamer2003}}
\tablenotetext{e}{\citet{gatewood2005}}
\label{alfoph_tab}
\end{deluxetable}

\clearpage
\begin{deluxetable}{lccc}
\tabletypesize{\scriptsize}
\tablecaption{Distinct Modes detected for $\alpha$~Oph.   
\label{table_modes}}
\tablewidth{0pt}
\tablehead{
\colhead{Mode} &
\colhead{Center Frequency} & \colhead{Half-Amplitude\tablenotemark{a}} & \colhead{Mode} \\
\colhead{\#} &
\colhead{(cycles/day)} & \colhead{(millimag)} & \colhead{Family\tablenotemark{b}}
}
\startdata
 1 &  1.768$\pm$0.017 & 0.188$\pm$0.035 & A \\
 2 &  1.835$\pm$0.017 & 0.135$\pm$0.035 & \\
 3 &  1.902$\pm$0.017 & 0.152$\pm$0.033 & \\
 4 &  3.428$\pm$0.017 & 0.080$\pm$0.024 & \\
 5 &  3.495$\pm$0.017 & 0.160$\pm$0.025 & A \\
 6 &  3.545$\pm$0.017 & 0.108$\pm$0.024 & \\
 7 &  3.695$\pm$0.017 & 0.268$\pm$0.024 & B \\
 8 &  5.439$\pm$0.017 & 0.140$\pm$0.028 & B \\
 9 &  6.923$\pm$0.017 & 0.175$\pm$0.029 & A \\
10 &  7.024$\pm$0.017 & 0.101$\pm$0.029 & \\
11 &  7.182$\pm$0.017 & 0.207$\pm$0.029 & B \\
12 &  8.075$\pm$0.017 & 0.183$\pm$0.033 & \\
13 &  8.508$\pm$0.017 & 0.292$\pm$0.030 & \\
14 &  8.617$\pm$0.017 & 0.115$\pm$0.028 & A \\
15 &  8.825$\pm$0.017 & 0.224$\pm$0.028 & \\
16 & 10.227$\pm$0.017 & 0.091$\pm$0.023 & \\
17 & 10.469$\pm$0.017 & 0.104$\pm$0.025 & \\
18 & 10.619$\pm$0.017 & 0.243$\pm$0.028 & \\
19 & 11.720$\pm$0.017 & 0.405$\pm$0.033 & \\
20 & 12.028$\pm$0.017 & 0.228$\pm$0.037 & A \\
21 & 12.412$\pm$0.017 & 0.312$\pm$0.039 & B \\
22 & 13.096$\pm$0.017 & 0.144$\pm$0.035 & \\
23 & 16.124$\pm$0.017 & 0.349$\pm$0.039 & \\
24 & 16.174$\pm$0.017 & 0.411$\pm$0.039 & \\
25 & 17.183$\pm$0.017 & 0.115$\pm$0.027 & \\
26 & 18.209$\pm$0.017 & 0.091$\pm$0.025 & \\
27 & 18.668$\pm$0.017 & 0.655$\pm$0.026 & \\
28 & 18.818$\pm$0.017 & 0.109$\pm$0.027 & \\
29 & 19.252$\pm$0.017 & 0.111$\pm$0.026 & \\
30 & 19.936$\pm$0.017 & 0.109$\pm$0.026 & \\
31 & 20.228$\pm$0.017 & 0.093$\pm$0.028 & \\
32 & 20.286$\pm$0.017 & 0.151$\pm$0.027 & \\
33 & 20.420$\pm$0.017 & 0.120$\pm$0.027 & \\
34 & 20.512$\pm$0.017 & 0.210$\pm$0.027 & \\
35 & 21.713$\pm$0.017 & 0.282$\pm$0.032 & \\
36 & 22.155$\pm$0.017 & 0.117$\pm$0.035 & \\
37 & 22.205$\pm$0.017 & 0.299$\pm$0.034 & \\
38 & 22.480$\pm$0.017 & 0.146$\pm$0.037 & \\
39 & 23.631$\pm$0.017 & 0.223$\pm$0.027 & \\
40 & 23.807$\pm$0.017 & 0.092$\pm$0.027 & \\
41 & 24.582$\pm$0.017 & 0.136$\pm$0.027 & \\
42 & 25.166$\pm$0.017 & 0.144$\pm$0.026 & \\
43 & 25.250$\pm$0.017 & 0.114$\pm$0.027 & \\
44 & 25.416$\pm$0.017 & 0.272$\pm$0.027 & \\
45 & 25.633$\pm$0.017 & 0.111$\pm$0.027 & \\
46 & 27.001$\pm$0.017 & 0.112$\pm$0.030 & \\
47 & 29.120$\pm$0.017 & 0.105$\pm$0.029 & \\
48 & 29.304$\pm$0.017 & 0.172$\pm$0.030 & \\
49 & 30.980$\pm$0.017 & 0.072$\pm$0.022 & \\
50 & 31.130$\pm$0.017 & 0.089$\pm$0.022 & \\
51 & 32.949$\pm$0.017 & 0.060$\pm$0.014 & \\
52 & 34.392$\pm$0.017 & 0.049$\pm$0.015 & \\
53 & 35.718$\pm$0.017 & 0.149$\pm$0.015 & \\
54 & 35.877$\pm$0.017 & 0.113$\pm$0.015 & \\
55 & 39.597$\pm$0.017 & 0.149$\pm$0.014 & \\
56 & 43.292$\pm$0.017 & 0.157$\pm$0.022 & \\
57 & 48.347$\pm$0.017 & 0.036$\pm$0.010 & 
\enddata
\tablenotetext{a}{Amplitude error derived from level of frequency-dependent 
noise background (see \S4.2).}
\tablenotetext{b}{See \S\ref{gmodes} for descriptions of g-mode families A and B.}
\end{deluxetable}

 \clearpage

\begin{figure}[p]
\begin{center}
\includegraphics[width=6in]{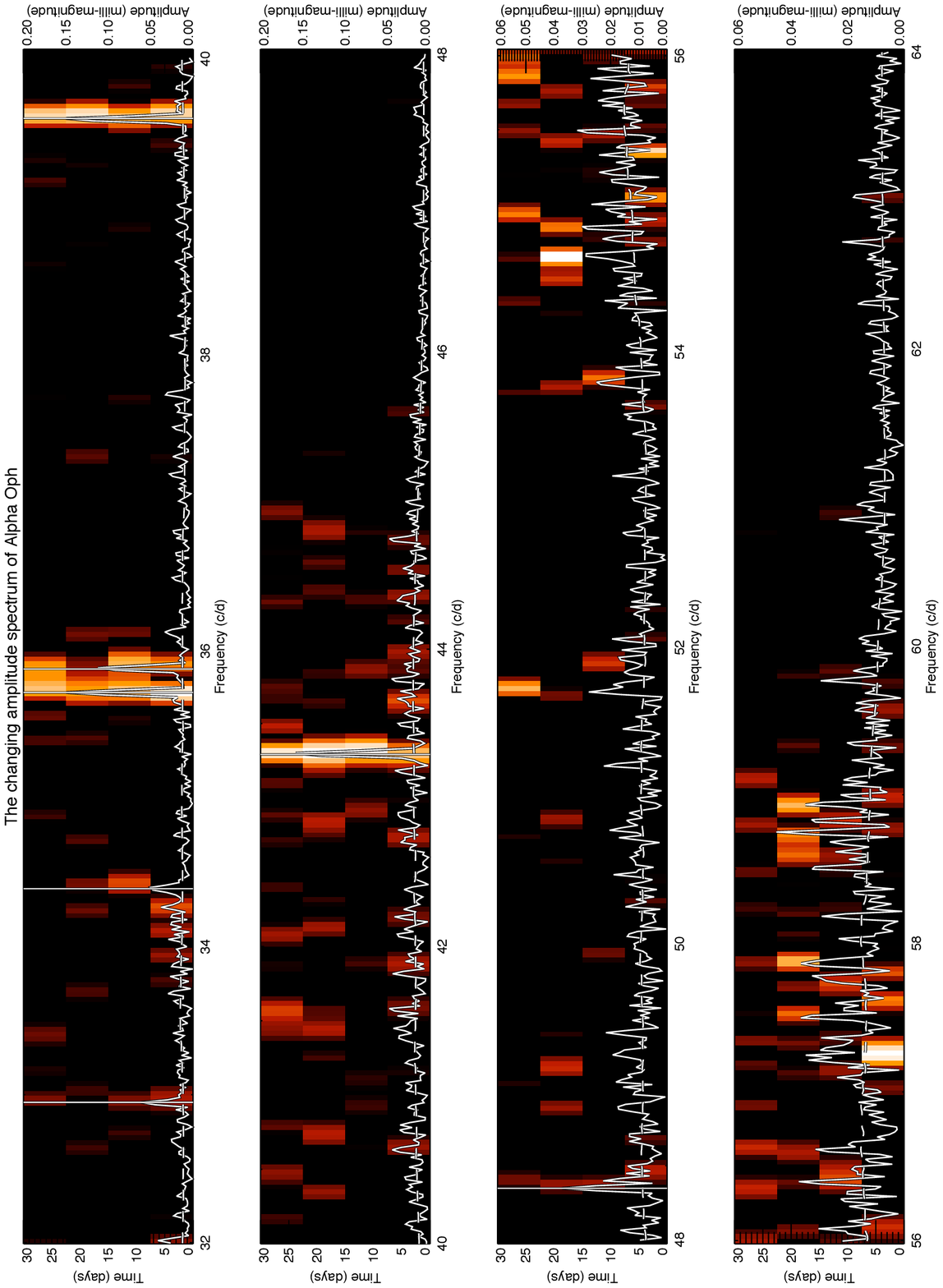}
{\\
Figure~\ref{fig3} (continued)}
\end{center}
\end{figure}

\begin{figure}[p]
\begin{center}
\includegraphics[angle=90,width=6in]{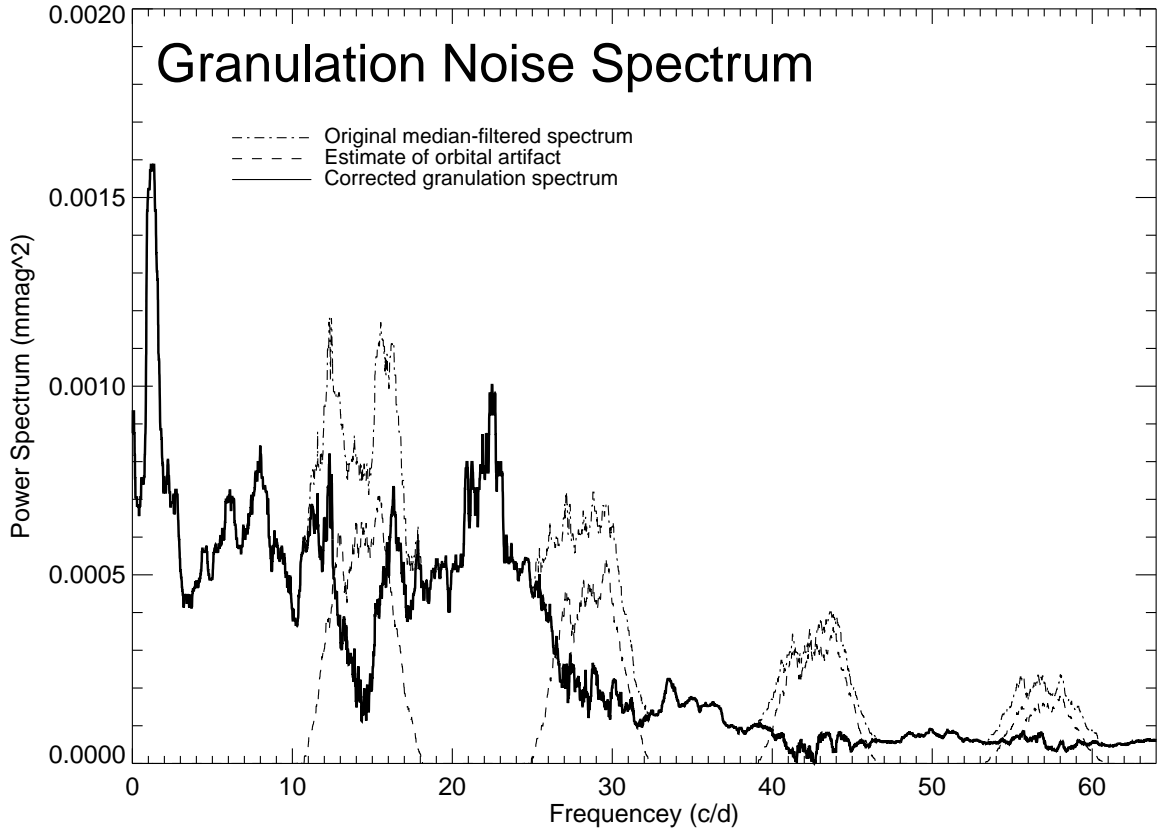}
\figcaption{\footnotesize 
This figure shows the median power spectrum for the light curve of $\alpha$~Oph using a 2~c/d window (dot-dashed line).  We note the orbital artifacts at harmonics of 14.2 c/d, and we have crudely  estimated the orbital artifact (dashed line).  
We also plot the corrected granulation spectrum (thick solid line), identifying a clear frequency break at 26$\pm$2~c/d.
\label{fig_noise}}
\end{center}
\end{figure}

\begin{figure}[p]
\begin{center}
\includegraphics[angle=90,width=6in]{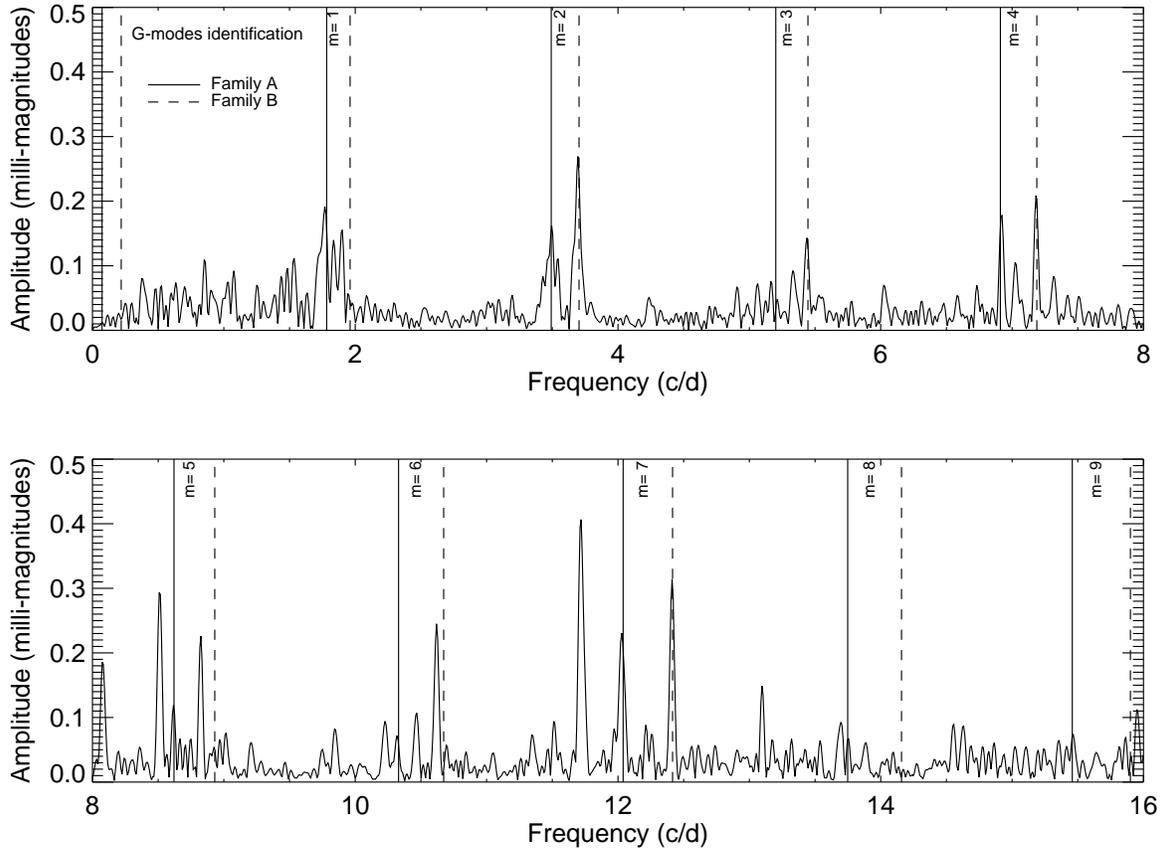}
\figcaption{\footnotesize
Here we repeat the  amplitude spectrum of the detrended light curve for frequencies below 16 c/d to highlight the g-modes.
We also include the identifications for 2 families of modes which are discussed in the text (\S\ref{gmodes}).
\label{fig_modes}}
\end{center}
\end{figure}

\end{document}